\documentclass[preprint2]{aastex}

\slugcomment{Not to appear in Nonlearned J., 45.}

\shorttitle{Microvariability of RLQs and RQQs}
\shortauthors{Ram\'\i rez, de Diego, Dultzin, Gonz\'alez-P\'erez}

\usepackage{graphicx}
\usepackage{subfigure}
\usepackage[subfigure]{ccaption}
\usepackage{color}

\begin{document}


\title{Multiband Comparative Study of Optical Microvariability in RL vs. RQ Quasars}


\author{A. Ram\'\i rez$^{1,2}$, J.A. de Diego$^1$, D. Dultzin$^1$, AND J.-N. Gonz\'alez-P\'erez$^3$}
\affil{1 Instituto de Astronom\'\i a, Universidad Nacional Aut\'onoma de M\'exico, Apartado Postal 70-264, 04510 M\'exico, D.F., M\'exico }
\affil{2 Instituto de Astrof\'\i sica de Andaluc\'\i a (CSIC), Apdo 18080 Granada, Spain}
\affil{3 Hamburger Sternwarte, Gojenbergsweg 112, 21029 Hamburg, Germany}
\email{aramirez@astroscu.unam.mx, jdo@astroscu.unam.mx, deborah@astroscu.unam.mx,jngonzalezperez@hs.uni-hamburg.de}




\begin{abstract}

We present the results of an optical multi-band (BVR) photometric
monitoring program of 22 core-dominated radio-loud quasars (CRLQs) and 22 radio-quiet quasars (RQQs). The aim was to compare the properties of microvariability in both types of quasars. We detected optical microvariability in 5 RQQs and 4 CRLQs. Our results
confirm that microvariability in RQQs may be as frequent as in CRLQs. In addition we compare microvariability duty cycles in different bands. Finally, the implications for
the origin of the microvariations are briefly discussed.

\end{abstract}





\section{INTRODUCTION}

Optical Microvariability (OM hereafter) defined as flux changes on
timescales ranging from minutes to hours, and with amplitudes that
range from a few hundredths to a few tenths of a magnitude, was
reported almost simultaneously as the quasar discovery \citep{Matthews63,Oke67}. However, those reports were ignored or
skeptically attributed to instrumental and/or weather causes.
Nowadays, with the CCD technology, and better statistical
techniques, there is no doubt about the reality of OM.

The importance of OM studies resides on the opportunity that they
provide to set limits on the size of the emitting regions, to
constrain emission models, and to probe physical conditions in the innermost
regions of active galactic nuclei (AGN), very close to the central
black hole. Some OM studies have been carried out to investigate its
incidence in radio quiet quasars (RQQs) as compared to radio loud
quasars (RLQs) \citep[][hereafter Paper I]{Jang97,Romero99,Stalin04,Gupta05,Stalin05,Carini07,de98}. The motivation for searching OM in RQQs and
compare its incidence with OM in RLQs, is to identify the source of
OM and try to constrain the most probable models: accretion disk
and/or jet emission \citep[e.g.][]{Lightman74,Blandford79,Marscher85,George91,Qian91,Marscher92,GK93a,GK93b,Mangalam93,Chak94,Krishan94}.

Some of these studies have found that OM occurs in quasars
regardless of their radio properties. The results reported by \citet{Stalin04} indicate that although strong radio emission does not
guarantee microvariability generation the most variable objects are
those with more intense radio emission. On the other hand, \citet{Carini07} found a higher instance of OM in radio-loud quasars than in radio quiet quasar. According to these results,
orientation could play a fundamental role. Thus, objects with jets
aligned closer to the line of sight are those with larger duty
cycles and amplitudes. On the other hand, in Paper I we showed for
the first time that  microvariability is a phenomenon as common in
radio quiet quasars (RQQs) as in core-dominated radio loud quasars
(CRLQ), thus indicating that OM in quasars is independent on the
radio properties, pointing to a discrepancy with respect to the
orientation scheme.

The orientation scheme assumes that OM is generated in the jet. Even
if only to test this  scheme it is relevant to investigate whether
OM does indeed origin in the jet, the accretion disk, or both. Disk
phenomena are mostly related with thermal processes such as hot
spots or other disk instabilities, while jet emission is associated
with non-thermal activity such as the relativistic particle ejection
events at the base of the jet. So in principle we can distinguish
the OM origin, if we can identify the OM event as the result of a
thermal or a non-thermal phenomenon. We will propose below a color
study that can help us  find some distinctive characteristic of the
variable component, which can be seen in the variations in the shape
of the continuum spectrum.

The paper is organized as follows: we begin in Section~\ref{datos} by
describing the observations, the data treatment, and the selection
criteria. In Section~\ref{res} we show the results for the detected
microvariations. In Section~\ref{discusion} we discusses these results,
and finally we summarize our work and give conclusions in
Section~\ref{summarize}.

\section{SAMPLE SELECTION, OBSERVATIONS AND DATA REDUCTION}\label{datos}


Because selection effects may yield unwanted biases in the study of
OM, we have minimized them using the same selection criteria
as in Paper I. Briefly, two quasar samples were observed and compared:
one constituted by 22 RQQs, and a control sample conformed by 22
RLQs (see Table \ref{lamuestra4}). The samples were chosen in such a
way that for each RLQ there is a RQQ with similar brightness and
redshift (differing less than $10\%$), avoiding possible evolution
biases between the samples. To avoid orientation biases, RLQs chosen
are all core-dominated RLQs (CRLQs), but avoiding to include objects extremely variables such as known blazars. When this sample was assembled, PKS~1510-089 was not considered a blazar, although now it is indeed. Unfortunately, it is hard to avoid the possibility to contaminate a sample of radio loud quasars by blazars, due to better/new observations, particularly in other frequencies. Nevertheless, we will show in a forthcoming paper \citep[]{Ramirez09} that the OM detected in this object presents characteristics that differ of those associates to blazars. For this reason, we prefer to keep it in our sample. Choosing CRLQs instead
of RLQs is a simple way to avoid unknown orientation effects in the
sample. Thus, the CRLQs are expected to be as homogeneous as
possible from the observational point of view, while their intrinsic
physical properties are a truthful representation of the RLQs
population. All the objects in the CRLQs sample have radio to
optical flux ratio near or above 1000, computed from the 408 MHz to
{\it V}-band, thus they belong to the RLQ population after \cite{Falcke96} criterion. Table \ref{lamuestra4} shows the complete sample. A
number identifies each couple (col. 1), ''a'' refers to RQQ
component and ''b'' to CRLQ; the object name (col. 2); the
coordinates (J2000.0) of each one (cols. 3 and 4); the reported
average magnitude in the {\it V} band (col. 5) and the
redshift (col. 6).


\begin{table}
\begin{center}
{\footnotesize \caption[The sample]{\bf The sample.
\label{lamuestra4}}
\medskip
\begin{tabular}{p{0.3cm} p{1.9cm} p{0.8cm} c p{0.3cm} c}
\hline \hline
\smallskip
Pair&              & R.A.     & DEC.        &      &        \\
ID  & object       & (J2000.0)& (J2000.0)   & V    & z      \\
\hline
1a  & US 1867      & 08:53:34 & +43:49:01   & 16.4 & 0.513   \\
1b  & 3C 334.0     & 16:20:21 & +17:36:29   & 16.4 & 0.555   \\
2a  & 1628.5+3808  & 16:30:13 & +37:58:21   & 16.8 & 1.461   \\
2b  & 3C 298       & 14:19:07 & +06:28:35   & 16.7 & 1.439   \\
3a  & 1222+023     & 12:25:17 & +02:06:56   & 17   & 2.05    \\
3b  & PKS~0421+019 & 04:24:08 & +02:04:30   & 17   & 2.048   \\
4a  & 1E 15498+203 & 15:52:02 & +20:14:02   & 16.5 & 0.25    \\
4b  & PKS 1217+02  & 12:20:11 & +02:03:42   & 16.5 & 0.24    \\
5a  & US 737       & 09:35:02 & +43:31:12   & 16.3 & 0.456   \\
5b  & PKS 1103-006 & 11:06:31 & -00:52:52   & 16.4 & 0.426   \\
6a  & 0214-033     & 02:17:29 & -03:08:08.6 & 16.8 & 0.323   \\
6b  & PKS 2208-137 & 22:11:24 & -13:28:09.7 & 17   & 0.39205 \\
7a  & CSO 233      & 09:39:35 & +36:40:01   & 17   & 2.03    \\
7b  & PKS 1022-102 & 10:24:56 & -10:31:44   & 17   & 2       \\
8a  & CSO~21       & 09:50:45 & +30:25:19   & 17   & 1.19    \\
8b  & PKS 1127-14  & 11:30:07 & -14:49:27   & 16.9 & 1.187   \\
9a & TON 156      & 13:21:16 & +28:47:19    & 16.6 & 0.549   \\
9b & PKS 1327-21  & 13:30:07 & -21:42:04    & 16.7 & 0.528   \\
10a & TON 133      & 12:51:00 & +30:25:42   & 17   & 0.56    \\
10b & 3C~281       & 13:07:53 & +06:42:13   & 17   & 0.599   \\
11a & CSO 18       & 09:46:36 & +32:39:51   & 17   & 1.3     \\
11b & PKS 0514-16  & 05:16:15 & -16:03:08   & 16.9 & 1.278   \\
12a & E 0111+388   & 01:13:54 & +39:07:45   & 16.7 & 0.234   \\
12b & PKS 2247+14  & 22:50:25 & +14:19:50   & 16.9 & 0.237   \\
13a & MRK 1014     & 01:59:50 & +00:23:41   & 15.6 & 0.163   \\
13b & PKS 2349-01  & 23:51:56 & -01:09:13   & 15.3 & 0.173   \\
14a & Q0050-253    & 00:52:44 & -25:06:51   & 16.1 & 0.626   \\
14b & PKS 2243-123 & 22:46:18 & -12:06:51   & 16.4 & 0.63    \\
15a & US 3150      & 02:46:51 & -00:59:31   & 16.8 & 0.467   \\
15b & PKS 0003+15  & 00:05:59 & +16:09:49   & 16.4 & 0.45    \\
16a & US 3472      & 02:59:38 & +00:37:36   & 16.6 & 0.532   \\
16b & PKS 0122-042 & 01:24:34 & -04:01:05   & 17   & 0.561   \\
17a & MRK~830      & 14:50:26 & +58:39:44.8 & 16   & 0.21    \\
17b & OX 169       & 21:43:35 & +17:43:49.5 & 15.7 & 0.213   \\
18a & PG 1444+407  & 14:46:46 & +40:35:06.0 & 15.6 & 0.267   \\
18b & PKS 2135-14  & 21:37:45 & -14:32:55.1 & 15.5 & 0.2     \\
19a & PG 1543+489  & 15:45:29 & +48:46:09   & 16.4 & 0.4     \\
19b & PKS 1510 -08 & 15:12:50 & -09:05:60.0 & 16.5 & 0.361   \\
20a & MC3~1750+175 & 17:52:46 & +17:34:20.3 & 15.5 & 0.507   \\
20b & PKS 2128-12  & 21:31:35 & -12:07:04.5 & 16.1 & 0.501   \\
21a & PG 1538+477  & 15:39:33 & +47:35:33   & 16   & 0.77    \\
21b & PKS 1424-11  & 14:27:37 & -12:03:54   & 16.4 & 0.805   \\
22a & PG 1407+265  & 14:09:23 & +26:18:18   & 15.9 & 0.944   \\
22b & PKS 2145+06  & 21:48:05 & +06:57:35   & 16.4 & 0.99    \\
\hline
\end{tabular}
}
\begin {list}{}
{\setlength {\rightmargin} {0.5truecm} \setlength {\leftmargin}
{0.2truecm} \setlength {\parsep} {0.2ex} \footnotesize}
\item [] Radio quiet quasars are identified with the letter a; radio loud quasars are identified with a letter b.
\vspace {-0.2truecm}
\end {list}
\end {center}
\end {table}

Observations were made in several seasons, between 1998 and 2002,
with four telescopes distributed in Mexico and Spain. In Mexico, two
telescopes, of 1.5-m and 2.1-m, were used (Mx1 and Mx2, hereafter,
respectively). These are operated by the Observatorio Astron\'omico
Nacional (OAN), located in San Pedro Martir, Baja California.
Observations in Spain were made using two telescopes. The 1-m
Jacobus Kapteyn Telescope (JKT), located in the Roque de los
Muchachos Observatory, and operated by the Isaac Newton group. We
also used the 1.5-m telescope at the Estaci\'on Observacional de
Calar Alto (EOCA), located in the Astronomical Hispanic-German
Centre in Calar Alto (CAHA), which is operated by the Spanish
Observatorio Astron\'omico Nacional.

Filters {\it BVR} of the Johnson-Cousins series were used throughout
all the observations. The detectors used were: in Mx1, a SITE SI003,
1024$\times$1024 pixels of 24-$\mu m^2$ which has a methacrome II
cover and a VisAr, to improve the response in the blue; in Mx2, a
Thomson TH7398 2048$\times$2048 pixels of 24-$\mu m^2$ which has a
methacrome II cover; in JKT, a TEK1024AR constituted by
1024$\times$1024 pixels of 24-$\mu m^2$ was used, this chip is
covered with Ar; and, in the EOCA telescope, a CCD Tektronics
TK1024AB of 1024$\times$1024 pixels of 24-$\mu m^2$. In all the
cases the detectors were binned to allow for a fast and low noise
readout.

The observational strategy was performed using
an \emph{Analysis of Variance} (ANOVA) design, similar to the one described in Paper
I, but considering three filters instead of only one. Shortly,
each time that an object was observed, it was recorded in five
observations of $\sim 1$-minute exposure each, in each filter in the
sequence {\it BVR}, except for a couple of cases, where the sequence
was inverted to {\it RVB}. It was possible to monitor from 2 to 4
pairs per night, and 2 to 5 times each pair. Each object was observed at moderate air masses, always at least 30 degrees above the horizon. The magnitude
difference errors were obtained directly from the observations by
means of the differential photometry. These errors range typically
from $\sim 0.001$ to $0.01$ instrumental magnitudes. Additionally,
each pair was observed on the same nights and in overlapping
sequences, to avoid possible biases due to atmospheric and/or
instrumental conditions.

Previous to the extraction of data, images were corrected for bias
and flat fields. The flat fields were sky flat-fields acquired at
the beginning and/or end of every night. Objects were observed,
insofar as possible, when they were meridian crossing, minimizing
color effects by the atmospheric extinction. For each object, field-stars were
used as comparison and reference stars to obtain differential photometric data used in the posterior analysis; typically, from four to eight stars were
used (except for a couple of cases where we only had two stars). The
selected stars were those that remained stable during the
observations at a $20\%$ confidence level. Here the results with
only two of them  will be shown. Unlike Paper I standard stars were observed to obtain the
flux level of the observed objects as well. These stars are all
taken from \citet{Landolt92}.

The IRAF/APPHOT\footnote{IRAF is distributed by the National Optical Astronomy Observatories, which are operated by the Association of Universities for Research in Astronomy, Inc., under cooperative agreement with the National Science Foundation.} package was used to perform data reduction. The
aperture radii used ranged from three to six arcseconds, while the size of the sky annuli was of 10 arseconds for the inner radius, and 16 arcseconds for the outer radius. When the host
galaxy was detectable, we took a sufficiently large aperture to
avoid effects due to the seeing. This was done although \citet{Carini91} and \citet{Clements01} found that the host galaxy
cannot provoke false microvariations due to light that {\it enters
and leaves} the aperture \citep[see also][]{Kidger92}.
However, Cellone, Romero \& Combi (2000) found that the host can
influence spurious variations detection, but only under extreme
atmospheric conditions (with FWHM variations of $\sim 5$ arcseconds)
for galaxies as bright as the nucleus. Similar conditions were
absent in our study.




\section{RESULTS}\label{res}

Searching for microvariability for each object was done with the ANOVA, particularly, the one-way ANOVA test was utilized. Contrary to a {\it standard
test}, where the object standard deviations are analyzed with respect to
comparison stars \citep[e.g., $\sigma_{obj} / \sigma_{est} = 2.576$, for
detection of variability,][]{Jang97,Romero99},
ANOVA is an empirical very robust statistical test. We used a
significance level of $0.1\%$ (corresponding to $3\sigma$) for the
one-way ANOVA test as variability criterion, although for some objects, in some filter, we will discuss also possible variations considering $3\%$ and $5\%$. The functional procedure on the ANOVA test is given in the next lines.

We have implemented One--Way ANOVA to search for variability comparing data measured in $k$ groups of observations sampled from a quasar lightcurve. If $y_{ij}$ represents the value of the $i$th observation ($i=1$, 2,...$n_j$; note that we have chosen $i=1$, 2, ...5 for all the groups) on the $j$th group ($j=1$, 2, ...$k$), we can express a mathematical model describing each single observation:

$y_{ij} = \overline{y} + g_j + \varepsilon_{ij}$

\noindent
where $\overline{y}$ is the mean of the whole dataset, $g_j$ is the between-groups deviation ($g_j = \bar{y}_j - \bar{y}$) and $\varepsilon_{ij}$ the within-groups deviation, often called residual or measurement error ($\varepsilon_{ij} = y_{ij} - \bar{y}_i$).

ANOVA tests whether the means of the groups are equal. Statistically, this can be expressed saying that the null hypothesis (i.e., the hypothesis that we are testing) is that the means of the different groups are the same, and the alternate hypothesis (i.e., the hypothesis that we will accept if the null hypothesis would be rejected) is that at least one group has a mean different from the others. In our case, the alternate hypothesis implies a microvariability detection.

Following the implicit mathematical model of ANOVA, the total sample variation can be split into variation \emph{between} and \emph{within} groups:

$$\displaystyle\sum_{j=1}^{k} \displaystyle\sum_{i=1}^{n_j} (y_{ij} - \bar{y})^2 =  \displaystyle\sum_{j=1}^{k} n_j (\bar{y}_j - \bar{y})^2 + \displaystyle\sum_{j=1}^{k} \displaystyle\sum_{i=1}^{n_j} (y_{ij} - \bar{y}_j)^2,$$

\noindent
where the left term is a measurement of the total deviations of the data with respect to the mean, the second term measures the total variation between groups, and the third term the total errors. The equation can be abbreviated to:
$$SS_{Total} = SS_{Group} + SS_{Error}.$$

When the null hypothesis is true, the $k$ groups sample data will be normally and independently distributed, with mean $\mu$ and variance $\sigma^2$. Thus the statistic:
$$F = {MS_{Group} \over MS_{Error}} = {SS_{Group} / (k-1) \over SS_{Error} / (N-k)},$$

\noindent
will follow an $F$ distribution with $k-1$ and $n-k$ degrees of freedom, and where the \emph{pseudo} variances $MS_{Group}$ and $MS_{Error}$ are mean estimates for the variations between groups and errors, respectively. Given a certain significance level $\alpha$, if the F statistic exceeds the critical value $F_{(k-1, n-k,\alpha)}$ the null hypothesis should be rejected.

The rest of this section is devoted to describe the results, object
by object, for those cases where variability was detected. In
Figures \ref{fig1} we show the light curves for differential
photometry for each object. Symbols represent the mean for each
group of five observations of each object (left panel) and the stars
(right panel). Error bars correspond to a standard error toward each
side, and they were calculated from the dispersion of each group.

From a total of 130 observations, we detected 9 microvariability events, and 2 possible ones, in eight
out of 44 observed objects, three CRLQs and five RQQs. Two quasars
displayed microvariability in more than one night: the CRLQ 3C~281 was observed to vary in two nights, while microvariability in Mrk~830 was detected one night, and probably were present in two more occasions.

\subsection{Radio Loud Quasars}\label{RLQ}


{\bf 3C~281.-} This radio loud quasar is known to reside in a very
rich galaxy cluster \citep{Yee87}. A variation of 0.15
magnitudes in one year has been reported in the infrared
\citep{Enya02a,Enya02b}. We observed this object for Paper I on May of
1997, but microvariability was not detected. We observed this quasar
with EOCA on 2000 March 1 and 4, detecting OM events in both
nights.


\begin{figure}
\centering
\subbottom[3C 281, 2000 March 1]{\label{fig1a}\includegraphics[scale=0.9]{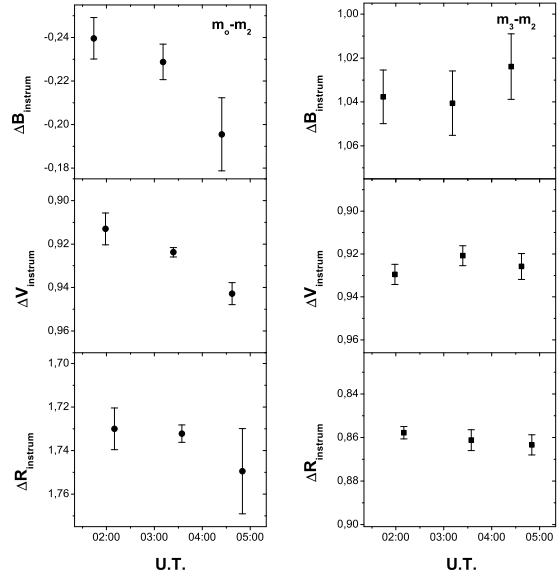}}
\subbottom[3C 281, 2000 March 4]{\label{fig1b}\includegraphics[scale=0.9]{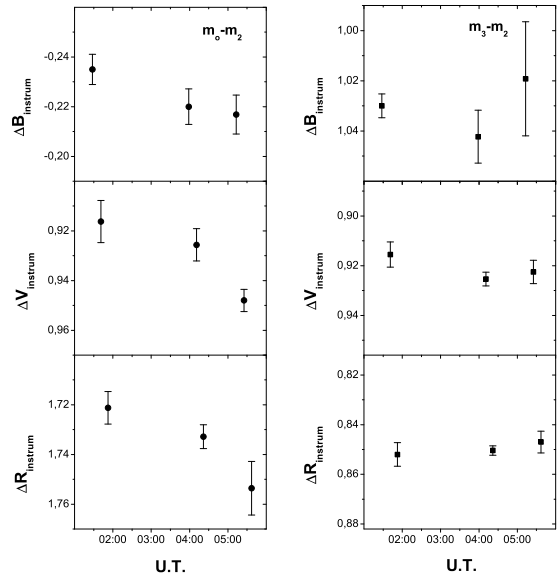}}
\caption{Light curves for the objects discussed in the text. In each Figure, left panel shows the data for the quasar, while right panel for the comparison stars. Symbols represent average values of five consecutive observations, and the error bars have been estimated from data dispersion. Footnote of each Figure indicates the quasar name and the observation date.}\label{fig1}
\end{figure}

{\it 2000 March 1.-} The object was observed during 2.7 hours three
times. A visual analysis of the light curves shows a decrease in
brightness of almost 4 hundredths of a magnitude (Figure
\ref{fig1a}). ANOVA confirms a microvariation in {\it B} and {\it
V}, detectable at a $0.1\%$ significance level, but not in {\it R}.
This detection is appreciated with all the available stars. In {\it
V}, the amplitude is $\Delta V= 0.031 \pm 0.007$ mags, in 2.7 hrs,
while between the second and third groups $\Delta V= 0.024 \pm
0.007$ mags, in $\sim$1 hour. In {\it B}, the amplitudes are $\Delta
B= 0.05 \pm 0.01$ mags between the first and the third
observations and $\Delta B= 0.03 \pm 0.01$ mags between the
second and the third.

{\it 2000 March 4.-} 3C~281 was observed during 3.75 hours in three
occasions. A visual revision of the light-curves establishes the
occurrence of a possible variation in the three bands (Figure
\ref{fig1b}). However, ANOVA only detected a variation in {\it V}
and a marginal one in {\it R} (with a significance level of 5\%).
The variation amplitude is $\Delta V= 0.027  \pm  0.006$ mags
between the first group and the third.

{\bf PKS~1510-089.-} This RLQ has shown intense activity on long and
intermediate-term variability \citep{Liller75,Tornikoski94,Villata97}, as well as correlated optical/radio
variability \citep{Tornikoski94}. OM also has been reported
\citep{Villata97,Xie01,Dai01,Xie04}. In particular, \citet{Dai01} reported a variation of 2
mag in 41 minutes in {\it R} band at the end of 2000 May. However,
\citet{Romero99} and \citet{Romero02} did not
detect any activity when they observed it between 1998 and 1999. Our
observations show that PKS~1510-089 was in a stage of activity
during the first semester of 2000. We monitored this object during
4.3 hours in 5 different occasions during the night of 2000 March 20 \citep[two months before the observations of][]{Dai01}.
Light-curves (Figure \ref{fig1c}) indicate the presence of an OM
event detected in {\it V} and {\it R}. While a progressive increase
in brightness is detected in {\it V}, a variation is observed
between the first and the second datasets in {\it R}, which remains
in the same level of brightness the rest of the session. ANOVA
agrees with the visual revision of Figure \ref{fig1c}. This
variation is confirmed with a $0.1\%$ level of significance. The
comparison between the stars indicates that they do not vary. In
{\it V}, the microvariation seems to evolve during the 4.3 h of
observation, with a total amplitude of $0.104 \pm 0.014$ mags (with
respect to the first data set changes are: $\Delta V= 0.03 \pm
0.02$ mags in 1 hour, $\Delta V= 0.07 \pm 0.01$ mags in 2.2
hours, and $\Delta V= 0.09 \pm 0.01$ mags in 3.25 hours). In {\it
R}, the amplitude between the first dataset and the average of the
subsequents is $\Delta R = 0.108 \pm 0.008$ mags. From the
light-curves, it is clear that color changes occurred during this OM (see Figure \ref{fig1c}).


\begin{figure}
\centering
\contsubbottom[PKS~1510-089]{\label{fig1c}\includegraphics[scale=0.9]{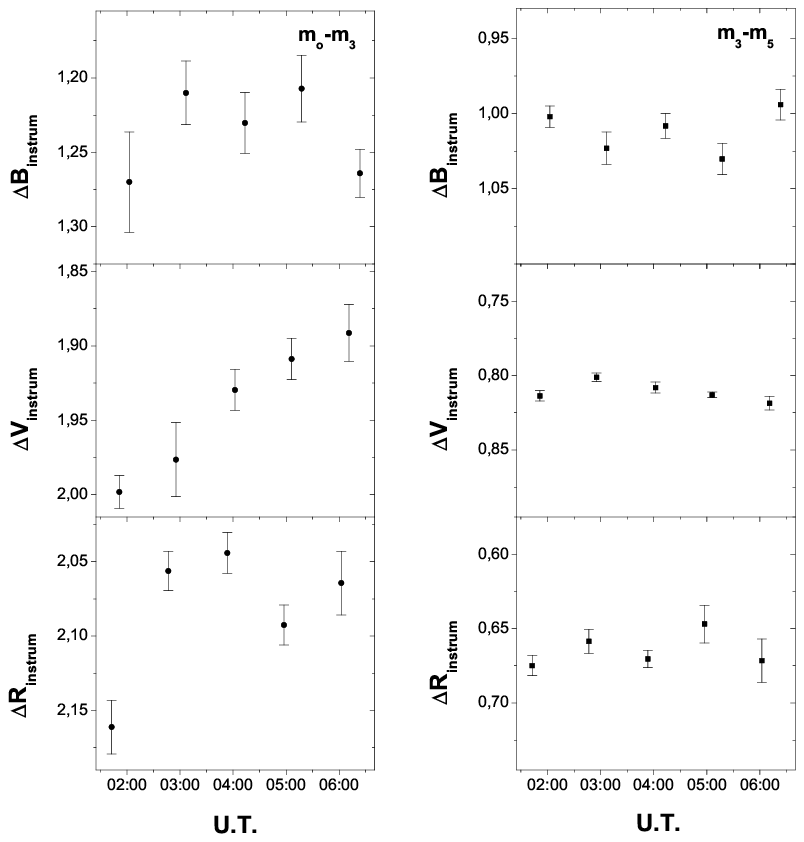}}
\contsubbottom[PKS~0003+15]{\label{fig1d}\includegraphics[scale=0.9]{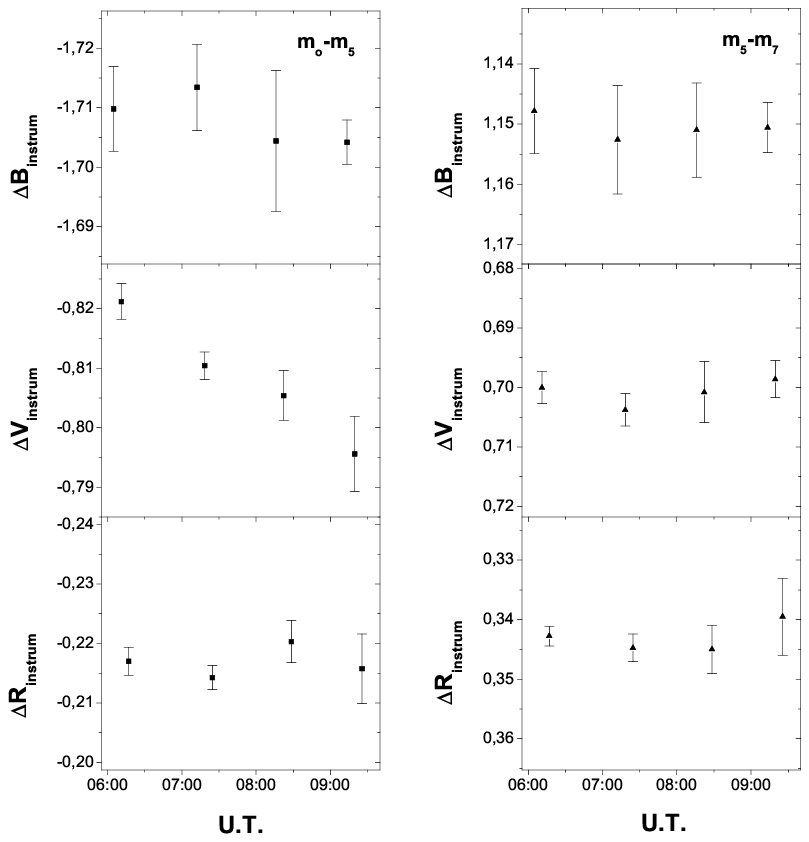}}
\contcaption{Continued.}
\end{figure}

{\bf PKS~0003+15.-} Historically, this quasar has shown intense
activity on long-term and intermediate-term variability \citep{Barbieri79,Pica88,Schramm94,Guibin95,Guibin98,Garcia99}. OM has also been detected.
On October of 1994, \citet{Jang95} observed a variation in {\it
R} with an amplitude of $0.11$ mags in 2 hours, while \citet{Eggers00} reported changes of $0.005$ mag/h in {\it R}. \citet{Garcia99} also reported microvariability for this quasar. PKS~0003+15
was reported in Paper I with no evidence of microvariability. We
have monitored this quasar during 3.13 hours in four occasions on
October 11 2001. Figure \ref{fig1d} shows that while brightness
remained constant in {\it B} and {\it R}, a decrement in {\it V} was
detected. ANOVA confirms this visual revision. The brightness for
this object showed a decrease of $\Delta V= 0.025 \pm 0.005$
mags during the monitoring time.





\subsection{Radio Quiet Quasars}\label{RQQ}

{\bf MC3~1750+175.-} The only variability event previously reported for this
quasar corresponds to radio wavelengths \citep{Fanti81,Mantovani82}. On 1999 August 20, we observed this object three
times during 3.13 hours. The light-curves show a microvariability
event only visible in the {\it V} band (Figure \ref{fig1e}). ANOVA
confirms this appreciation, the amplitude of this variation is $
\Delta V = 0.04  \pm 0.008$ mags between the first dataset and the
third.


\begin {figure}
\centering
\contsubbottom[MC3 1750+175]{\label{fig1e}\includegraphics[scale=0.9]{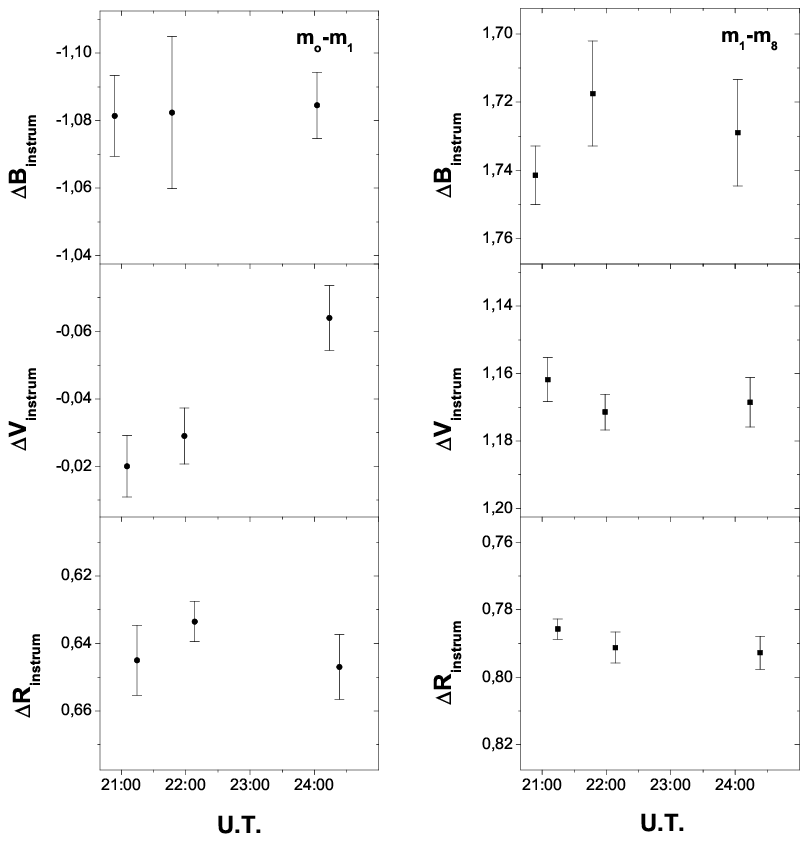}}
\contsubbottom[CSO 21]{\label{fig1f}\includegraphics[scale=0.9]{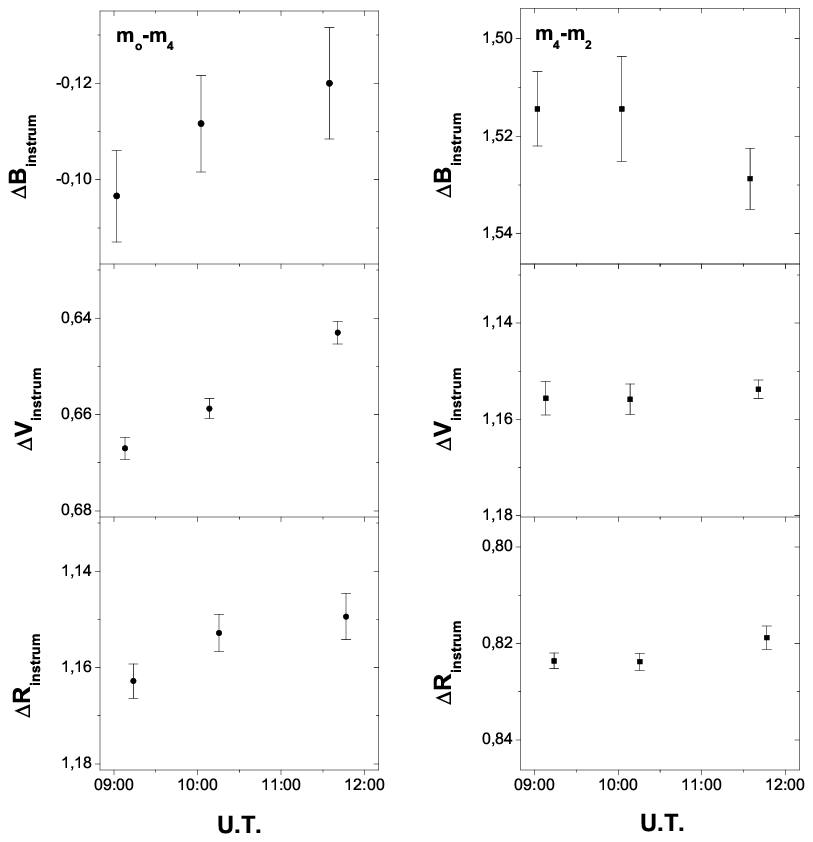}}
 \contcaption{Continued.}
\end{figure}

{\bf CSO~21.-} Observations for CSO~21 were reported in Paper I, but
without a positive detection of microvariability. For the present
study, the object was observed in three occasions on 2001 December, during 2.5 hours. A gradual brightness increase is appreciated
in the three bands (Figure \ref{fig1f}). However, ANOVA  indicates a
variation in {\it V} only, a barely marginal variation in {\it R} (with a
significance level of $10\%$), and no variation in {\it B} (if the
observed tendency in the {\it B} light-curve is a variation, it is
hidden by the internal data dispersion). The amplitude of the
variation is $\Delta V= 0.025 \pm 0.005$ mags in 2.5 hours.
Variation amplitude of the {\it R} band is $ \Delta  R=0 .016  \pm
0.004$ mags between the first dataset and the third.

{\bf 1628.5+3808.-} In Paper I we reported the cases of microvariability for this object for the first time. They are two events on 1997 February an May. In February, a variation with amplitude of $0.082$ mags
throughout 2.4 hours; while in May, throughout six monitoring hours,
the OM consisted of pulsations with amplitudes from $0.04$ to
$0.057$ mags. We observed this quasar in two occasions on 2001 June 13. The light curves show an appreciable brightness decrease in
the three bands (Figure \ref{fig1g}). ANOVA indicates that this
variation is detectable in the {\it B} and {\it R}, but in {\it V}
the variation is only marginal (the significance level is  $3\%$).
The data dispersion in the comparison stars do not show evidence for statistical differences. The OM amplitudes are $\Delta  B = 0.061 \pm
0.011$ mags and $\Delta  R = 0.018  \pm  0.004$ mags, in the 2.39
hours of monitoring. The {\it V} marginal detection
would have an amplitude of $0.03 \pm 0.01$ mags.


\begin {figure}
\centering
\contsubbottom[1628.5+3808]{\label{fig1g}\includegraphics[scale=0.9]{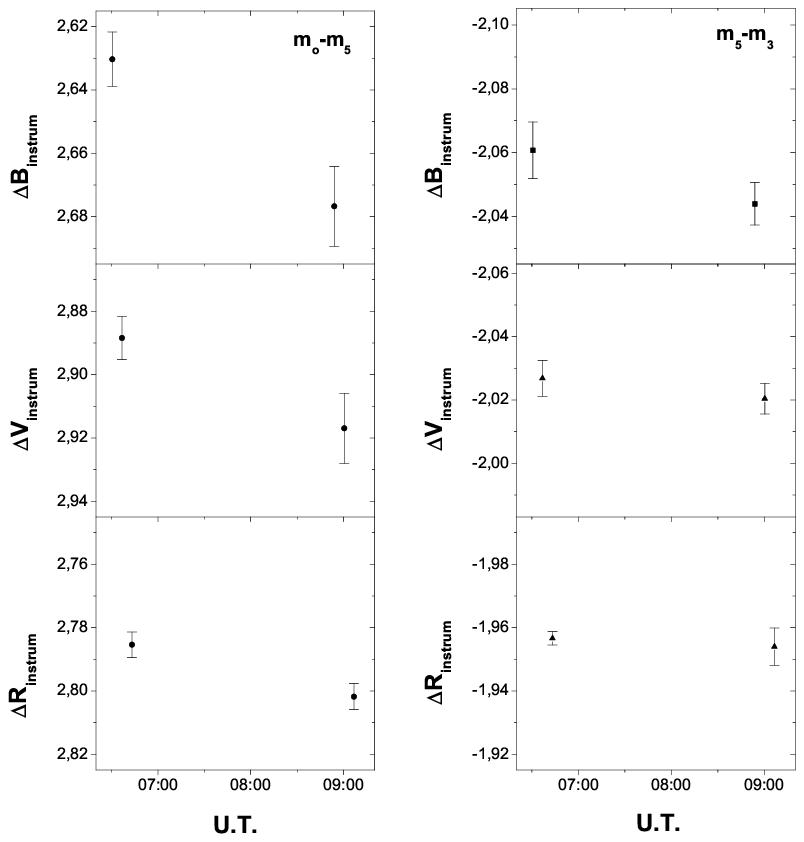}}
\contsubbottom[US 3472]{\label{fig1h}\includegraphics[scale=0.9]{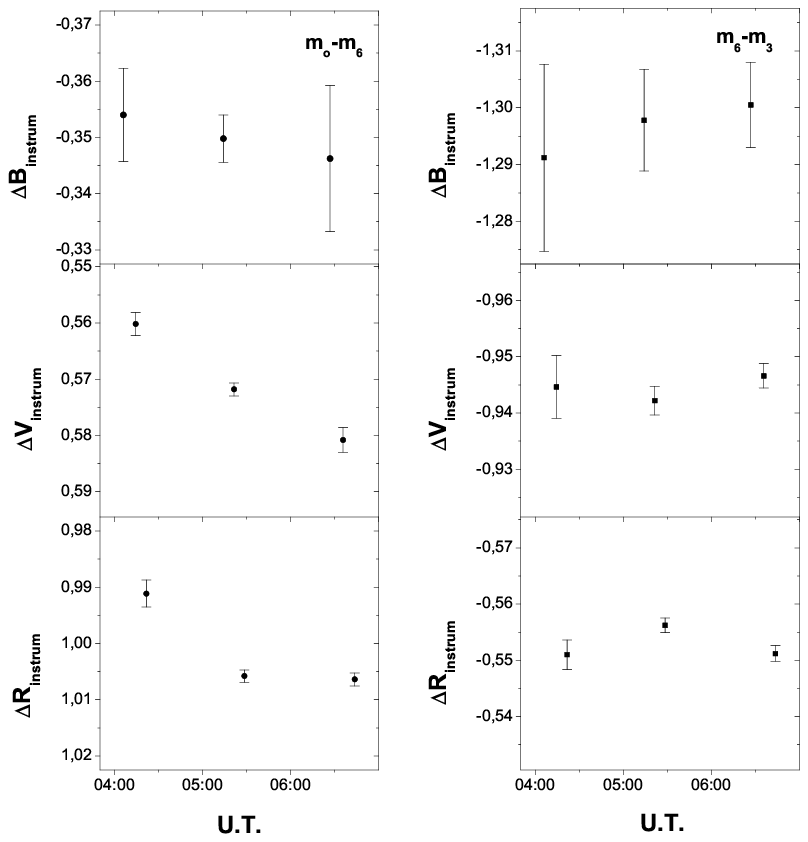}}
\contcaption{Continued.}
\end{figure}

{\bf US~3472.-} In Paper I we reported an OM event for this quasar
with an increase of $0.02$ mags in 50 minutes. During 2 hours, a
series of oscillations followed to this brightness increase, with
amplitudes between $0.040$ and $0.055$ mags. Long-term optical and
infrared variability  has also been reported \citep{Enya02a}. We
observed this object during 2.5 hours in three occasions on 2001 December
20. The light-curves show a brightness decrease in {\it V} and
{\it R} (Figure \ref{fig1h}). ANOVA confirms the variation. In {\it
V}, the microvariation seems to have kept a constant rate with a
total amplitude of $\Delta V=0.020 \pm 0.003$ mags in 2.35 hours.
Between the first and the third dataset in the {\it R} band, the
brightness varied by $0.014 \pm 0.002$ mags.

{\bf Mrk~830.-} Long- and short-term variability, and microvaribility have been reported for this
quasar. We observed this object in four different occasions, OM
events being detected on 1999 August 17 and 18, and on 2001 June 14.
The object possessed a brightness very similar on the three
occasions, $V \sim 17.6$. This brightness is similar to that
reported by \citet{Stepa01} ($V=17 .29 \pm 0.03$), and
{\citet{Chavu95} ($V=17 .62$). Unfortunately, the fields of
the two telescopes used are different, and thus it was more
convenient to use a different set of comparison stars for the August of
1999 and the June of 2001 observations.

{\it 1999 August 17.-} The object was observed during 2.13 hours in
three times. A visual revision of the light-curves indicates the
possible detection of a microvariability event in {\it V} (Figure
\ref{fig1i}). Due to instrumental problems, in the first sequence of the {\it R} band, we missed the object for several images. Thus we can present only the complete curve for the comparison stars. ANOVA shows that this variation would be
marginal, with a significance level of $3\%$. The increment in
brightness, if real, has a constant rate with a total amplitude of
$\Delta V= 0.03 \pm 0.01$ mags in 2.31 hours.


\begin {figure}
\centering
\contsubbottom[Mrk 830, August 17 1999]{\label{fig1i}\includegraphics[scale=0.9]{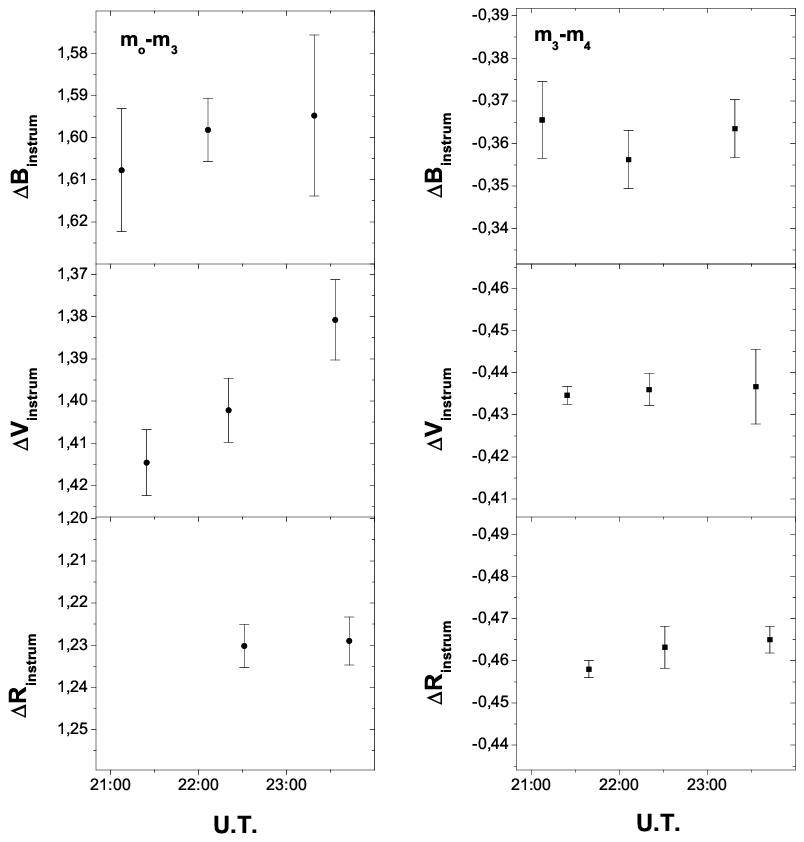}}
\contsubbottom[Mrk 830, August 18 1999]{\label{fig1j}\includegraphics[scale=0.9]{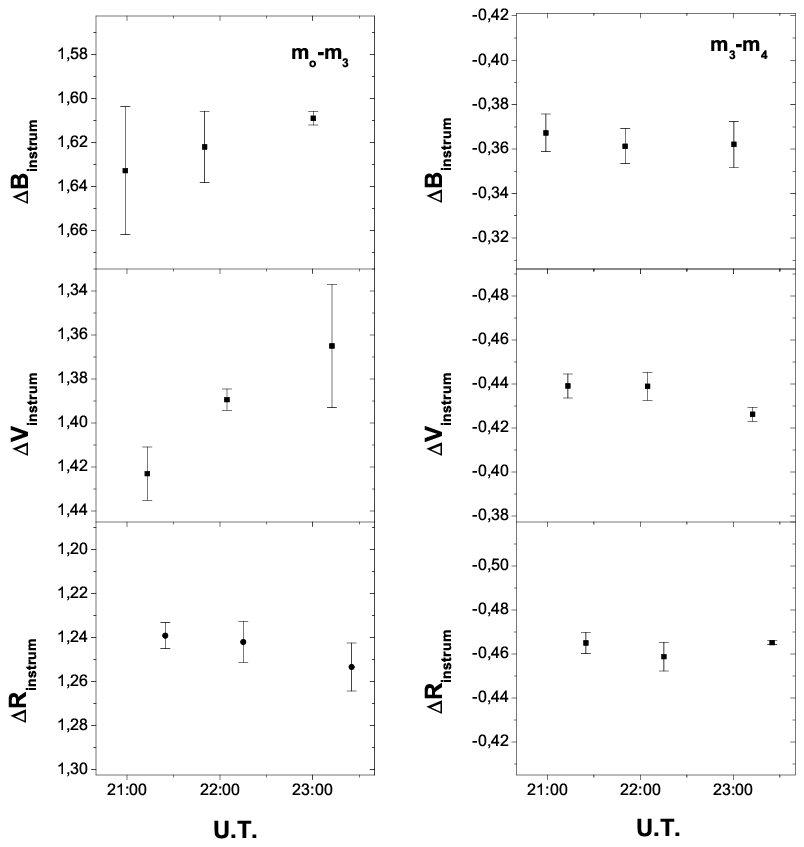}}
\contcaption{Continued.}
\end{figure}

{\it 1999 August 18.-} The monitoring dataset consisted on three
observation groups taken during two hours. Again, a revision of the
light-curves shows a possible variation in {\it V} (Figure
\ref{fig1j}). ANOVA detects this feature as a marginal variation, with a
significance level of $3\%$. As on the previous night, the tendency
is a variation  at a constant rate, but in this time with a total
amplitude of $\Delta V= 0.05 \pm 0.03$ mags.

{\it 2001 June 14.-} On this night the quasar was observed twice. A
visual revision of the light-curves indicates a possible variation
in the three bands (Figure \ref{fig1k}). The brightness  increase is
a few hundredth of magnitude between the two groups of observations.
ANOVA confirms the reliability of this variation. While in {\it V}
and {\it R} the significance level is $0.1\%$, in {\it B} the value is
$ \sim $ 5\%. The amplitudes are $ \Delta V = 0.039 \pm 0.007$ mags
and $ \Delta R = 0.030 \pm 0.007$ mags, occurring during 1.18 hours.
If a variation exists in {\it B}, the amplitude would be of $ \Delta
B \sim 0.04$ mags.


\begin {figure}[t]
\centering
\contsubbottom[Mrk 830, June 14 2001]{\label{fig1k}\includegraphics[scale=0.88]{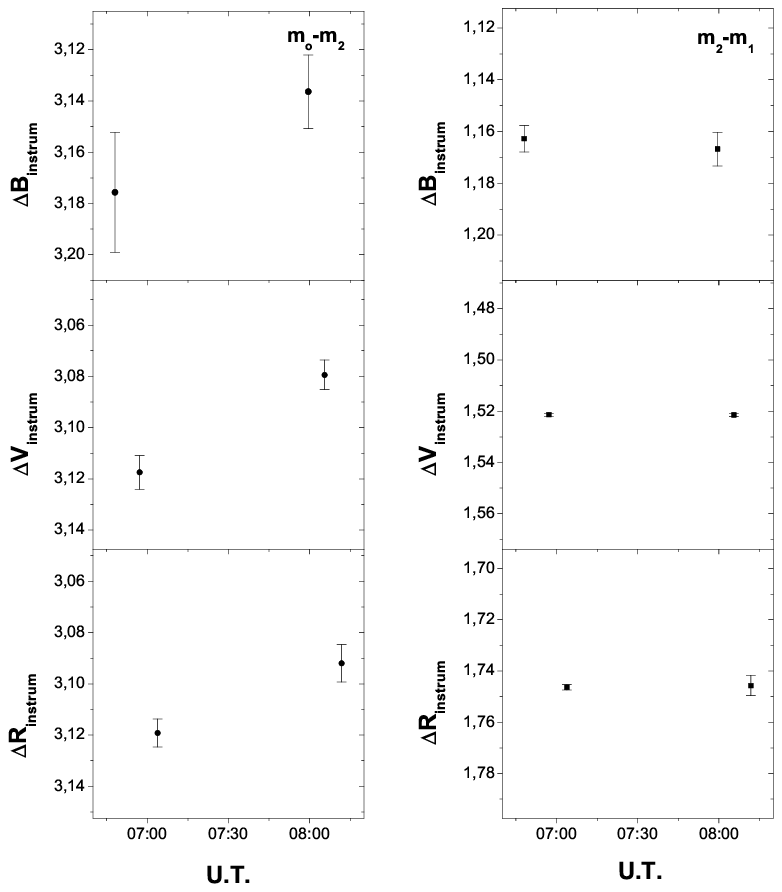}}
\contcaption{Continued.}
\subconcluded
\end{figure}

\section{DISCUSSION}\label{discusion}

An immediate result of this work is that microvariability is a more
complex phenomenon than appreciable in single optical band studies.
The different color evolutions during microvariability events will
be analyzed in a forthcoming paper \citep{Ramirez09}. In
this section we will compare statistically the results of OM
obtained from the two samples of RQQs and CRLQs. The aims are
twofold: 1) to find inferences on the populations of which the
samples have been extracted when comparing each pair of objects, and
2) to describe the probability of detecting a microvariability event
as a function of the duration of the observation for each object.
Several well-known statistical  tests are applied to attain these
objectives. Some of these tests deal with the qualitative
properties of the results, such as detection or non-detection of OM
(\S~\ref{chi-2}), and others deal directly with the numerical
results (\S~\ref{even}). Because of the differences in the
variability detected for each band, the numeric tests were carried
out separately.

Statistical tests are described in detail in Paper I. Shortly, a $
\chi ^ 2 $ test for homogeneity is used to compare the occurrence of
OM events between the CRLQs and the RQQs samples. Using this test, one can
determine if a quasar type presents a statistically larger number of OM events
than the other. On the other hand, a quantitative test on the
differences among the samples is carried out through the comparison
of the variances of the paired objects. The observed variance was
considered to arise from two different error sources: one
corresponding to the observational error and another due to the
intrinsic variance of each object (see Paper I for details). Then a
couple of statistical tests were applied to the mean of the
differences of intrinsic variances of RQQs and CRLQs: a {\it
t-Student} matched pair test and an independent sample test. At the
end of the section, the microvariability occurrence as a function of
monitoring time, i.e., the duty cycle will be analyzed.

The variations showed in Figs. \ref{fig1c} and \ref{fig1d}, for PKS~1510-089 and PKS~0003+15, respectively, differ very much in different bands. However, these variations can be explained in the context of emission models. We will explain such variations in a forthcoming paper \citep[]{Ramirez09}.

\subsection{$\chi^2$ Test for Homogeneity}\label{chi-2}

The $ \chi^2$ test for homogeneity is used to compare two or more
qualitative properties of several samples when the number of
elements in each sample is fixed. In our case, we will split the
observations of the two samples of RQQs and CRLQs into those that
show variability and those that do not.

For each sample of  RQ and CRL objects the microvariability events
were counted regardless of the band of detection.  As in Paper I we note that
although each observation will be considered independently, some of
them may belong to the same object. In doing this, some unwanted
effects can be introduced because the observations of the same
object may not be completely independent. This may be the case, for
example, if one object is intrinsically much more variable than
another.


\begin{table}
\begin{center}
{\footnotesize \caption{$\chi^2$ contingencies table.\label{tbl3}}
\begin{tabular*}{\columnwidth}{cccccc}
\hline\hline
 & \multicolumn{2}{c}{\underline {Observed}}  &  & \multicolumn{2}{c}{\underline{Prospective}} \\
Quasars & V & NV & Total & V & NV   \\
\hline
RQQ   & 5 & 58  & 63  & 4.5 & 58.5         \\
CRLQ  & 4 & 59  & 63  & 4.5 & 58.5         \\
Total & 9 & 117 & 126 & 9   & 117          \\
\hline
\end{tabular*}
}
\end{center}
{\footnotesize Contingencies Table to compare the radio properties
of the microvariability of the two quasars samples. It does not seem
to have any difference when the three filters are considered. This
result indicates that optical microvariability is a common
phenomenon to the object classes. $\alpha = 0.01$ as variability
criterion. }
\end{table}

The total number of paired observations for each sample is 63
(although the number differs for each band, i.e., the occasions in
which one could observe the two members of the same pair were 61 for
{\it B}, 63 for {\it V} and 61 for {\it R}). In 9 of these
monitoring microvariability was detected for at least one member of
each pair, with a significance level of $0.1\%$. Five out of these 9
were detected in RQQs and four in RLQs. This number rises to 11 when
we take into account the two marginal detections of Mrk~830.

Table \ref{tbl3} shows the $\chi^2$ contingencies  for these data.
Column 1 displays the number of OM events with a significance level
of up to $0.1\%$. The number of observations in which there are no
evidences of variability appear in column 2, the total number of
observations appear in column 3, and the expected frequencies of
microvariations under the null hypothesis, i.e., if both types of
objects possess the same microvariability properties, are shown in
columns 4 and 5. These last columns are calculated multiplying the
partial totals of each line and each column and then dividing by the
total number of observations. The homogeneity test shows that the
probability of observing the numbers in Table \ref{tbl3} can be
obtained $27\%$ of the times when extracting two samples in an
aleatory way from the same population. The probability to obtain the
observed OM frequencies is large enough and thus the null hypothesis
cannot be rejected. In other words, it is reliable to assume
similarity between the two samples.

\subsection{Quantitative Test for Differences between RQQs and CRLQs}\label{even}

Although the test for homogeneity did not show differences for the
occurrences of OM between RQQs and CRLQs, a test on the quantitative
values of the variance for both samples of objects might show
whether the OM is larger in one type than in the other. The color
changes during variations make it interesting to consider
observations of each band  separately.

In Paper I we showed that it is better to estimate the difference
between RQQs and CRLQs breaking the variance into two components
than considering a single observed variance. The intrinsic variance
for a source, $ V_r $, can be written as $ V_r=V_o-e ^ 2/5 $, where
$ V_o $ refers to the observed variance and $e^2$ are the squared errors of the data weighted by the number of
observations in each group (five). The last refers to the
inherent error of the observation and reduction data processes, and
it is calculated empirically from the variance for each of
the five observation groups.

The {\it t-Student} test can be used  to find differences between
the means when these have a normal distribution. This is
approximately the case for the intrinsic variances difference. For
these data, we have computed the {\it matched pair test}, in which
the mean of the distribution of these difference should not be
significantly different from zero if both samples are extracted from
the same population. The test indicates that there are no
significant differences among the samples: it gives a probability
that the observed differences occur by chance of $18.7\%$ for {\it
B}, $19.1\%$ for {\it V} and, $87\%$ for {\it R}. When an
independent pair test is carried out, the mean difference values are
similar, within the errors, and so are the uncertainties of such
differences.

 This test  was repeated considering average values of the intrinsic
variances for each pair of objects. The qualitative result remains
the same,  however quantitative results  do change: in {\it V}, the
probability that the observed differences happen by chance increases
to $28\%$, in {\it B} changes to $23\%$ and in {\it R} it diminishes
to $53\%$. The averages in this case are distributed so that the
fits are quite good, because although we have lost \emph{degrees} of
freedom, the data of the variations themselves are masked when they
are averaged with other data. It is necessary to note that the
probability obtained for the data of the {\it V} band is consistent
with the value obtained in Paper I, where the same filter was used.
These results imply that the incidence of OM is independent from
luminosity and redshift.


\begin{figure}[t]
\centering
\subfigure[]{\label{fig4.7}\includegraphics[width=\columnwidth]{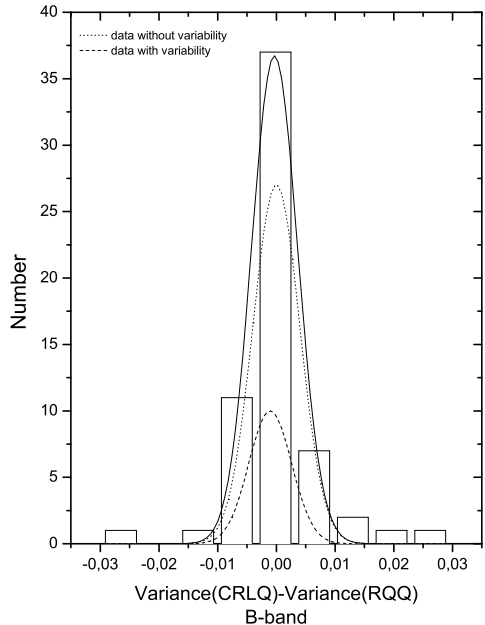}}
\caption{Histogram for the intrinsic variance differences among CRLQ and RQQ paired. The continuous line represents the result of adjusting the mixture of two normal distributions to the data without detected microvariability (dotted line) and those with detected microvariability at least in a single band (dashed line).}
\end{figure}

\begin{figure}[t]
\centering
\contsubfigure[]{\label{fig4.8}\includegraphics[width=\columnwidth]{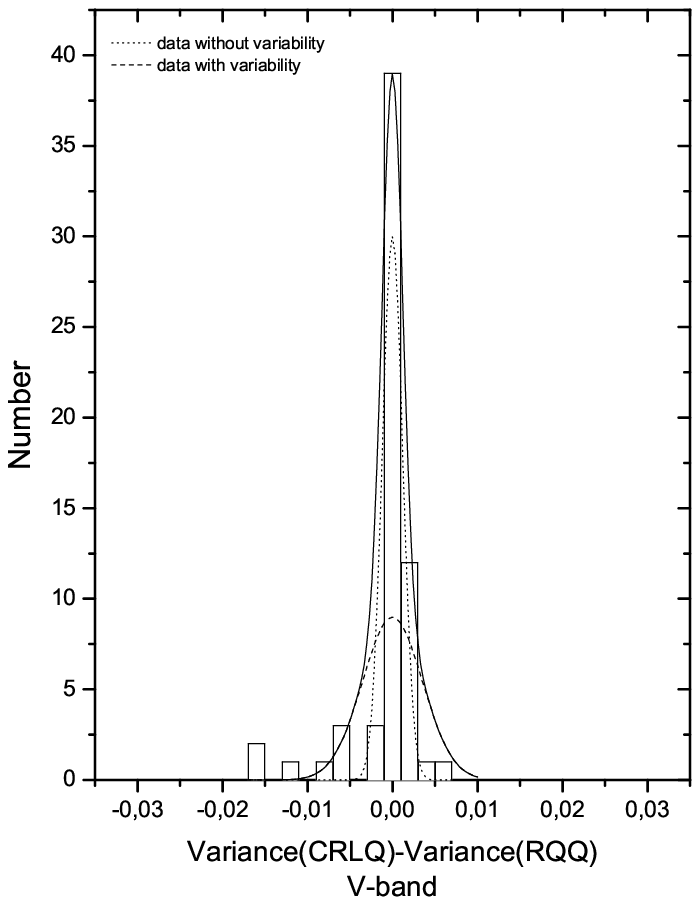}}
\contcaption{Continued.}
\end{figure}

\begin{figure}[t]
\centering
\contsubfigure[]{\label{fig4.9}\includegraphics[width=\columnwidth]{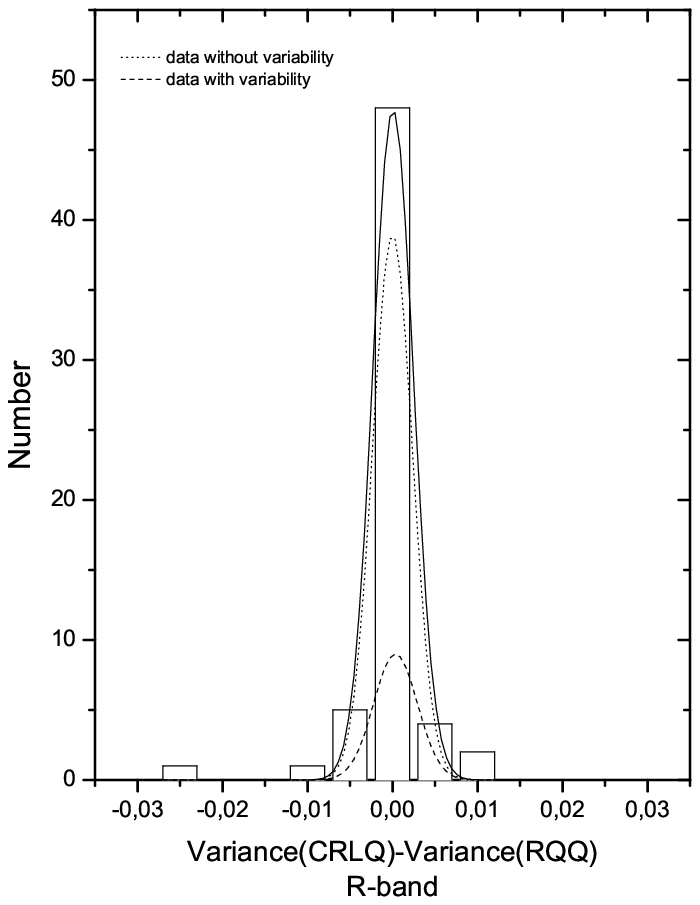}}
\contcaption{Continued.}
\subconcluded
\end{figure}

The tests neatly point out to an extremely small and non significant difference in the
microvariability displayed by the two samples, yielding no evidence that they come from different populations. Figures
\ref{fig4.7}, \ref{fig4.8} and \ref{fig4.9} show the histograms for
the intrinsic variance differences of the observations paired for
each filter.  Except in the case of {\it R}, for which data can be fitted using a single Gaussian profile with standard deviation of $2.3 \times 10^{-3}$ and an average value of $\sim 0$, the others cannot be
accurately fitted by a single Gaussian. Taking into account the
results presented in Paper I, we separated the data into
observations with (dashed line) and without (dotted line) detected
variations. Then, good fits are obtained. As in Paper I we found
that some variations could occur and not been detected. An OM
event could occur in the three bands even if it is detected in only
one.

In the {\it B} band, the data group where OM was not detected is
centered in $\sim0$, and it has a standard deviation of $4 \times 10^{-3} $, i.e., indistinguishable from zero (Figure \ref{fig4.7})
while, the group with variability has a standard deviation of $4
\times 10^{-3}$ and is centered at $1 \times 10^{-3} $. These
results indicate that the samples are not statistically
distinguishable. In {\it V} there is a similar behavior (Figure
\ref{fig4.8}). The non variable component was fitted with a Gaussian
that has a standard deviation of $1.2 \times 10^{-3} $ and is
centered on $4 \times 10^{-4} $, while the variable component is
centered at $ \sim 0$ and has a standard deviation of $4 \times 10^{-3} $. The data with OM detection were fitted by a
Gaussian centered on $4 \times 10^{-4} $ with standard deviation $3
\times 10^{-3} $, while the non detection data were best fitted
with a Gaussian centered on $0$ with standard deviation $2.3 \times
10^{-3} $ (Figure \ref{fig4.9}). These results can change when we
consider that some quasars were observed in more occasions than
others.

All the statistics discussed here indicate that there are no
significant differences in the OM properties between CRLQ and RQQ.
While this is at odds with some other studies, it is consistent with
the results reported in Paper I, by \citet{Stalin04}, and by
\citet{Gupta05}.

\subsection{Sample Selection and Duty Cycle}\label{DC4}

We can characterize the probability with which an OM event is
detected  as a function of the duration of the monitoring. The Duty
Cycle (DC) is a very useful tool to compare the behavior of
different types of AGN, and to plan future observations. The DC is
defined as the ratio of the number of observations with a positive
detection of OM  to the total number of observations, weighted by
the period of time during which the objects have been observed \citep[see][]
{Romero99}. Of course OM depends also on intrinsic causes
that determine its occurrence. The last however, are unpredictable
and unquantifiable. OM is believed to be a transient phenomenon.
Independently of its origin, neither thermal nor non-thermal
instabilities are expected to be permanent. We have calculated the
DC using  expression (2) from \citet{Romero99}, but considering
the joined observations for RQQs and CRLQs, because we detect no
difference between both samples. When the DC is calculated without
taking into account the band, a value of $8.45\%$ is obtained. If
marginal detections are considered, the value does not changes too
much: $10\%$. Separating data by band, we obtain $DC_B =2.3\%$;
$DC_V =6.8\%$, and $DC_R =5.8\%$.

The $DC_V$ is three times lower than the one reported in Paper I,
where the same filter was used. The discrepancy could be due to the
use of three filters in the present study. When we analyze the light
curves from Paper I, we can see that one third of the events
correspond to variations with constant increase or decrease. The
rest correspond to changes with oscillating behavior. If this
proportion is a general behavior, and since we have now a lower
temporal covering, we expect to detect only the {\it constant}
changes, i.e., a third of all the possible detections with one
filter.

For the blazar class, \citet{Sagar04} reported a $DC=72\%$, and
\citet{Gupta05} reported $DC=60\%$, for microvariability;
while, \citet{Carini90} reported a $DC$ of $\sim 80\%$ for timescales
from hours to days. For the quasars \citet{Stalin04} calculated
a $DC_{RQQ}= 17\%$, and $DC_{RLQ}=15\%$. Their observations were
carried out in the {\it R} band. Using the data from table 4 of
Paper I, and the expression of \citet{Romero99}, a DC of $18\%$
is obtained for the join of RQQs and CRLQs samples.

 Other groups have found differences in the Duty Cycle of
these objects. \citet{Romero99} calculated a DC of $68\%$ and
$6.9\%$ for RL and RQ objects, respectively (their observations were
carried out with a {\it V} filter). Taking the data of \citet{Jang97}, the DC$_ {RQQ}= 7.6\%$ (their observations were
carried out with a {\it R} filter), while DC$_{RLQ} = 60.6\%$. With
the data from \citet{Gk00}, DC$_{RQQ} =6.1\%$ (not
taking into account their detections that at least are {\it very
probable}, while when including those {\it probable detections}
DC$_{RQQ} = 29.4\%$). These results show a clear difference between
radio loud and radio quiet quasars, which represents a behavior
very different to the one found in the present study.

At this point we want to stress the relevance of a good sample
selection. However, all these groups included BL Lac objects among
the RLQ samples, and it is very possible that this can alter the
value of the DC for the RL sample. The DC for the RQQs on the other
hand are very similar ($\sim 6-7\%$) to that found by us (for the
complete sample). To illustrate this point, we point out that \citet{Romero99} use a sample of RL objects that includes several BL
Lac objects. 53\% of their observations and 67\% of their
microvariability detections, correspond to blazar type objects. On
the other hand, the RQQs sample of \citet{Jang97} is
constituted mainly by non OM objects, since it is conformed by 90\%
of low brightness quasars. The average magnitudes for RQs are
similar to the average magnitudes for RLs (averages only differ in
2\%), but the average redshifts are much smaller for RQs (average
redshifts differ in 62\%; for RQQ $z$ average is 0.5 while $z$
average for RLQ is 1.3). Although in Paper I we have not found
evidence of  dependence of OM on brightness and/or redshift, it
would be necessary to investigate further such possible biases. The
non homogeneous samples´ problem could also be present in the work
of \citet{Gk00}. These authors report observations of
RQQ, without a control sample, comparing with their previous results
on radio loud objects.

It is also worth noticing that, when the above authors make a
sub-selection of objects with the selection criteria described here
and in Paper I (except for the use of lobe dominated radio
quasars instead of CRLQ), they obtain similar comparative results to
those reported in Paper I \citep[see also][]{Stalin04}.

Finally, in Paper I we developed an alternative method to calculate
the Duty Cycle. This method considers that all the quasars present
the same DC in H hours of monitoring. We grouped the data in parcels
according to monitoring duration, with intervals of 1 hour. Parcels
larger than 6 hours have been excluded from the sampling since they
were very few for a good statistics. Using the current data in the
equations 1, 2, and 3 from Paper I, we get that the mean value for
the probability of detecting an OM event in 1 hour of monitoring,
$P_1$ has a value of $8.9 \pm 2\%$, which is marginally different from the
value found in Paper I ($P_1=5\pm 2\%$, however take also account the fact commented above that in this study the detection of oscillating variability may be dumped by the larger lags in the monitoring). Table \ref{tbl4} is similar
to Table 4 in Paper I. Columns show, respectively, the values of the
number of monitoring hours, $H$; the number of monitors in H hours,
$N_H$; the number of OM events detected in $H$ hours, $E_H$; the
probability of detect an OM event in $H$ monitoring hours, $P_H$;
and the probability of detection in 1 monitoring hour, $P_1$.


\begin{table}
\begin{center}
\caption{Duty Cycle\label{tbl4}}
\begin{tabular}{lcccc}
\hline\hline
H & $N_H$ & $E_H$ & $P_H$ & $P_1$ \\
\hline
1 &  9 & 1 & 0.111 & 0.111  \\
2 & 21 & 1 & 0.048 & 0.024   \\
3 & 53 & 5 & 0.094 & 0.032   \\
4 & 25 & 2 & 0.080 & 0.021   \\
5 & 17 & 0 & .....   & .....   \\
\hline
\end{tabular}
\end{center}
\end{table}

\subsection{ANOVA and $\chi^2$ Tests}\label{ANchi}

To illustrate the reliability in the use of ANOVA, additionally we have applied a test $\chi^2$ to the data.  The average, $B_i$, has been taken from each group of five observations, and the error, $s_i$, is obtained from the dispersion of these five observations. Table \ref{ANOVA1} shows the differential photometric data for each group of observations ($t_1$,$t_2$ and $t_3$ of US\,3472). Taking the whole data set mean, $<B>$, we can obtain the statistical $$\chi^2 = \sum {(B_i - <B>)^2 \over s_i^2}$$.

As example, we see the case of US 3472 for the V band. Please note that we are carrying a large number of decimal places during the calculations. For this quasar we have the following values obtained from differential photometry, comparing with a stars field
$$B_1= 0.5602 \pm 0.00206,$$
$$B_2= 0.5718 \pm 0.00116,$$
$$B_3= 0.5808 \pm 0.0022.$$

From these values, the median is $<B> = 0.57093$. Thus, we have $\chi^2 = 27.13095 + 0.5625 + 20.12746 = 47.82091$. As the tabulated value for $\chi^2$, with 2 degrees of freedom, for a level of meaning to the $0.1\%$ has a critical value of $13.8150$, we have detected variations at this level.

For the case of stars field, the raw differential photometric data is shown in Table~\ref{ANOVA2}, and the means are:
$$B_1= -0.9446 \pm 0.00558,$$
$$B_2= -0.9422 \pm 0.00256,$$
$$B_3= -0.9466 \pm 0.00216,$$
which median is $<B>= -0.94447$. With this, we have $\chi^2= 0.000542773 + 0.78627 + 0.97242 = 1.75923$.

For comparison, using ANOVA the sum of the differences of the averages of the data shown in Table~\ref{ANOVA1} is $0.00107$, while the square difference of the internal dispersion is $0.0002084$. As the degrees of freedom are $k-1=2$ and $N-k=12$, we have $F=30.80614203$. With a level of significance of $0.001$, a value of $12.97$ is obtained. With such result, we have a positive detection of OM.

\begin{table}
\begin{center}
\caption{US 3472 minus star 3 data\label{ANOVA1}}
\begin{tabular}{ccc}
\hline\hline
$t_1$ & $t_2$ & $t_3$ \\
\hline
0.558 & 0.571 & 0.583 \\
0.564 & 0.570 & 0.585 \\
0.555 & 0.569 & 0.585 \\
0.558 & 0.575 & 0.576 \\
0.566 & 0.574 & 0.575 \\
\hline
\end{tabular}
\end{center}
\end{table}

With the same procedure for the field stars (Table~\ref{ANOVA2}), we obtain that the sum of the square difference of the averages is $0.0000485335$, while the difference of the internal dispersion is $0.0008472$. Thus, we have that $F=0.34721671$.

\begin{table}
\begin{center}
\caption{US 3472: star 3 minus star 6 data\label{ANOVA2}}
\begin{tabular}{ccc}
\hline\hline
$t_1$ & $t_2$ & $t_3$ \\
\hline
0.952 & 0.944 & 0.948 \\
0.959 & 0.933 & 0.939 \\
0.936 & 0.948 & 0.951 \\
0.928 & 0.945 & 0.95 \\
0.948 & 0.941 & 0.945 \\
\hline
\end{tabular}
\end{center}
\end{table}

This example illustrates that microvariability detections with ANOVA are at
least as reliable as those that can be obtained using $\chi^2$. In our
example, we could estimate errors for each observation in the $\chi^2$ test
from the same quasar dataset, because the observations were performed using
an ANOVA design. This is not usually the case with error estimates in
$\chi^2$ based designs. In $\chi^2$ tests, errors are usually calculated
either from different data sets (one or more comparison stars) \citep[e.g.,][]{Romero99} or from IRAF
theoretical errors multiplied by a correction factor \citep[e.g.,][]{Stalin05}. Besides, this
estimated error is often considered as a constant along the dataset, missing
the internal variability of the data due to short timescales factors (tiny
clouds, seeing changes, etc.). By contrast, ANOVA uses
internally consistent errors calculated from the same dataset.

\section{SUMMARY AND CONCLUSIONS}\label{summarize}

We present a  comparative study of the occurrence of OM in radio
loud and radio quiet quasars using BVR bands. Following Paper I we
stress the importance of sample selection criteria and observational
methodology.  Microvariability events were analyzed using {\it
one-way} ANOVA test, reinforcing with a visual revision of the light-curves, and a $\chi^2$ test.
Five events were found in RQQs and four in CRLQs, with a
significance level of $0.1\%$ for a positive detection. Then,
microvariability properties for each sample were compared using a
$\chi^2$ test for homogeneity. No significant difference among the
samples was found, in agreement with the results obtained in Paper
I.

Additionally, it was shown that both quasar types present the same
quantitative microvariability properties. This is evidenced applying a
{\it t-Student} test to the variance differences of each CRLQ-RQQ
paired couple. Since quasars were paired by brightness and
redshift, these properties do not influence the result. Moreover,
these results do not differ when applying a statistical {\it
t-Student} test for independence on the total observations,
confirming the results reported in Paper I, i.e., that the evolution
of the objects seems to have little influence on microvariability.

Similar results have been found for long-term and intermediate-term
variability \citep[e.g.][]{Webb00,Vanden04}.
The union of all these results, and the discovery of radio jets in
RQQs \citep{Blundell98,Blundell01,Blundell03} suggest that the OM source must be the
same in both types of quasars, or at least very similar. This suggestion is reinforced by the detection of variations in radio frequencies for RQQ \citep[e.g.,][]{Fanti81, Mantovani82, Blundell03}.

Finally the Duty Cycle (DC; the percentage of monitoring time that
an AGN shows variability) was calculated. This probability is 2.3\%
in the {\it B} band, 6.8\% in the {\it V} band, and 5.8\% in the
{\it R} band. Without separating data by filter, DC= 8.45\% is
obtained. These differences of DC between bands might arise from spectral variability during the MO events.

Results obtained by \citet{Jang95,Jang97}, \citet{Gk00}, and \citet{Romero99} indicate that microvariability
could be related to radio emission properties of the quasars.
However, these results are probably biased due to sample selection
criteria (see 4.3). On the other hand, it has been shown that with
selection criteria similar to those discussed in Paper I, \cite{Stalin04} and \cite{Gupta05} obtain compatible results with those of the present paper.

As our results seem to indicate that the optical microvariability does not depend on the radio-properties of quasars, we would not exclude the possibility that it can have a similar origin in both RLQs and RQQs (at least for the commonly defined ranges). Although this would be an important result, deriving from comparative studies, we would still have the additional problem to establish unambiguously the origin of the microvariability. The possibility of a symbiosis disk-jet phenomenology \citep[e.g.,][]{Falcke96} can complicate any analysis, since instabilities (that produce variability) originated in the inner disk can be propagated to the jet \citep[e.g.,][]{Gupta05,Wiita05}.

In a forthcoming paper \citep{Ramirez09}, we shall
address this issue using our multiband data. Events arising in the
accretion disk are expected to show a thermal signature, while jet
perturbation events should be dominated by non-thermal processes.
Thus the spectral evolution during a change of brightness, in other
words, a color variability analysis can allow us to differentiate
thermal from non-thermal origins.




\acknowledgments

We want to acknowledge the careful reading of our manuscript by an anonymous referee and his comments, which helped to improve this paper. We thank Gabriel Garcia, Salvador Monrroy, Felipe Montalvo, Gustavo Melgoza, Michael Richer, Gaghik Tovmassian, Sergei Jarikov and the staff from the OAN for assistance during the observations. DD is grateful for support from grant IN100507 from PAPIIT-DGAPA, UNAM. JAD and AR are grateful for support from CONACyT grants 50296 and 149972, respectively. AR is grateful for support from CONACyT grant with number of request 000000000081535. Authors are also grateful for computational support to DGEP-DGAPA, UNAM.

\end{document}